\documentclass[9pt,twocolumn]{article}%[12pt, letterpaper]{article} # 
\usepackage[utf8]{inputenc}
\usepackage[T1]{fontenc}
\usepackage{comment}
\usepackage[english]{babel}
\usepackage{hyperref}
\usepackage{amssymb}
\usepackage{graphicx}
\usepackage{subfig} % for subfloat
\usepackage{amsmath}
\usepackage{xcolor} % for inkscape exported figures
\usepackage{pgfbaseimage}
\usepackage{hhline} % for table
\usepackage{booktabs} % for toprule, etc.
\usepackage{geometry}
\usepackage{enumitem}

\usepackage{datetime}

\usepackage{algorithm}
\usepackage[noend]{algpseudocode}
\usepackage{etoolbox}
\makeatletter
\patchcmd{\ALG@doentity}{\item[]\nointerlineskip}{}{}{} % fix the issue with different line height in algorithms with noend
\makeatother

\newgeometry{top=3.6cm,bottom=3.7cm}

\title{
  \vspace{-10ex}
  Pattern matching algorithms in Blockchain for network fees reduction
}

\author{
  Robert Susik$^{\dag}$, Robert Nowotniak$^{\ddag}$\\
  \\
  \parbox{\textwidth}{
    \centering
    \small \dag~{Lodz University of Technology, Institute of Applied Computer Science\\ Al.\ Politechniki 11, 90--924 Łódź, Poland}\\
    \href{mailto:rsusik@kis.p.lodz.pl}{rsusik@kis.p.lodz.pl}\\
    \small \ddag~{MetaSolid.tech, Al. Grunwaldzka 56/202, 80-241 Gdańsk, Poland},\\
    \href{mailto:rnowotniak@metasolid.tech}{rnowotniak@metasolid.tech}\\
    \vspace*{0.4em}
  }
}

\newdateformat{monthyeardate}{%
  \monthname[\THEMONTH], \THEYEAR}

\date{\monthyeardate\today}

\begin{document}

\twocolumn[
  \begin{@twocolumnfalse}
    \maketitle
    \begin{abstract}
      Blockchain received a vast amount of attention in recent years and is still growing. 
      The second generation of blockchain, such as Ethereum, 
      allows execution of almost any program in Ethereum Virtual Machine (EVM),
      making it a global protocol for distributed applications. 
      The code deployment and each operation performed in EVM cost the network fee called gas,
      which price varies and can be significant.
      That is why code optimization and well-chosen algorithms
      are crucial in programming on the blockchain. 
      This paper evaluates the gas usage of several exact pattern matching algorithms on the Ethereum Virtual Machine. 
      We also propose an efficient implementation of the algorithms in the Solidity/YUL language. 
      We evaluate the gas fees of all the algorithms for different parameters 
      (such as pattern length, alphabet size, and text size).
      We show a significant gas fee and execution time reduction with up to 22-fold lower gas usage and 55-fold speed-up comparing to StringUtils 
      (a popular Solidity string library).
      % In addition, we analyse the price of code execution in Ethereum network 
      % using Ganache platform.
       \vspace{3ex}

       \noindent\textbf{keywords: } blockchain, pattern, matching, ethereum, string
       \vspace{6ex}

    \end{abstract}
  \end{@twocolumnfalse}
]

\section{Introduction} \label{sec:intro}

\subsection{Background}
\label{subsec:background}

Blockchain emerged as a peer-to-peer network 
with immutable transaction records on a shared public ledger
designed for implementing transactions of electronic cash (cryptocurrency).
Nakamoto introduced the first successful implementation
in 2008 called Bitcoin~\cite{nakamoto2008bitcoin}. 
Over the years, blockchain gained popularity~\cite{gad2022emerging}
and became a promising technology that found 
% many applications.
application in many computer science fields.
Several alternative blockchains were
introduced (Namecoin\footnote{\href{https://www.namecoin.org}{www.namecoin.org}}, 
            Litecoin\footnote{\href{https://litecoin.org}{litecoin.org}}, 
            Peercoin\footnote{\href{https://www.peercoin.net}{www.peercoin.net}}, etc.)
% ~\cite{namecoin2010, litecoin2011, peercoin2012} 
before the second generation of blockchain was developed.

Ethereum~\cite{Buterin2013}, the first Blockchain 2.0,
was introduced as a protocol for building decentralized applications running in the blockchain.
% Ethereum can also be defined as 
In short, 
it is a distributed data storage plus smart contracts platform~\cite{guo2022survey}
that introduces an Ethereum Virtual Machine (EVM).
% While Bitcoin allow developers to write a scripts 
% for transactions, and it supports a large number of opcodes,
% it is very constrained in terms of programming.
% The point is that the Bitcoin's language is simple, stack-based,
% and not a Turing-complete~\cite{turing1936computable},
% that means, in short, it is missing loops,
% so it does not allow to implement 
% complex smart contracts in an efficient way.
% Ethereum introduces an Ethereum Virtual Machine (EVM),
One of the main advantages of EVM is the support of Turing-complete~\cite{turing1936computable}
programming language, 
which allows for writing decentralized applications based on smart contracts~\cite{szabo1996smart}.
% which can implement any logic.
Ethereum has its own cryptocurrency called Ether, 
which is also used as a computational crypto-fuel to execute 
a code in EVM and pay transaction fees.
For each transaction, 
the user needs to specify the upper bound of gas that can be consumed by the transaction.
An advantage of such an approach is 
that it helps to avoid the situation where 
all the user's resources are wasted in, 
for instance, an ``infinite'' loop.
The mentioned code execution cost may differ
depending on the number of operations performed
in a transaction. 
It means the infinite loop is not possible because, 
in the worst case,
the EVM will stop processing the code by raising the ``out of gas'' error.
% It is required to provide a gas limit when running 
% transaction to avoid a situation where all the caller
% resources are wasted in the "infinite" loop.
A single computational step
(which we can compare to a single CPU cycle)
costs one unit of gas,
and a single operation usually takes more than one step.
For instance, an operation (ADD) that sums two 32-byte integer numbers costs three units of gas.
On the contrary,
there are a few operations that cost nothing, such as RETURN. 
Apart from the execution cost, 
% there is a fee for 
each byte of the transaction
data costs 5 units of gas.

Several languages are available for writing 
smart contracts, such as Solidity, YUL, Serpent, and Vyper.
The former, Solidity, is the most popular~\cite{wu2019empirical} 
and recommended object-oriented programming language for Ethereum.
YUL and Serpent are high-level assembly languages,
and Vyper puts emphasis on simplicity and security
(the syntax is similar to Python, with inheritance removed).
All of the mentioned languages are translated to EVM stack-based bytecode language
that, once deployed in blockchain, can be executed by 
a transaction transferring Ether (the fuel) to the contract address.
% The code is executed by a miner, and the miner gets the payment 
% for code execution.

Development of blockchain high-level programming languages
opened new opportunities to create more complex smart contracts, 
which combined with user interfaces form applications called Dapps.
Dapps are aligned with the web3 concept 
where the applications are decentralized
and always available.
However, 
more complex apps consume more gas which causes higher costs.

\subsection{Motivation}
\label{subsec:motivation}

Blockchain finds a wide range of applications
in areas such as 
healthcare~\cite{kumar2020novel, adere2022blockchain},
voting~\cite{sober2021voting},
transportation~\cite{mollah2020blockchain},
music industry~\cite{li2021decentralized},
supply chains~\cite{zhang2022blockchain},
reputation systems~\cite{almasoud2020smart},
document versioning~\cite{nizamuddin2019decentralized},
and decentralized finance~\cite{amler2021defi}.
The interest in Blockchain-based technologies is growing rapidly~\cite{wu2019empirical, belchior2021survey}.
% which is aligned with web3 development.
Similarly to Web2.0, web3 Dapps
can be reached with its alias name via DNS-like services
% such as~\cite{ensdomains, unstoppabledomains, namecoin2010}
such as ENS\footnote{\href{https://ens.domains}{ens.domains}}, 
Unstoppable domains\footnote{\href{https://unstoppabledomains.com}{unstoppabledomains.com}}, or 
Namecoin
that are supported by web browsers 
(web browser extensions or dedicated web browsers to navigate and browse
blockchain-based applications).
This new type of application has 
web/mobile apps, 
backend (smart contract) 
data sources (Oracle contract),
or data storage (i.e., IPFS).
% and allowing developers to create 
% applications that have 
% web interface,
% mobile app, 
% and backend as a smart contract. 

In a general case, it is possible to implement almost any 
application with the use of blockchain technology.
However, there are technical (i.e., stack depth and size)
and financial (gas is expensive compared to CPU time) limitations.
While the first one may be solved 
with future Ethereum Virtual Machine (EVM) development,
the gas fees seem to be a challenge~\cite{marchesi2020design, mars2021machine, kim2021predicting}.
Currently, 
there are approaches that significantly reduce the cost of running code in blockchain, 
such as Polygon, Solana, or Ethereum 2
(a new ``consensus layer'', 
which leverages \textit{Proof-Of-Stake} algorithm~\cite{king2012ppcoin} 
in place of \textit{Proof-Of-Work}~\cite{nakamoto2008bitcoin}).

The transformation process of traditional applications 
to a Dapp is not well defined and is a subject of study~\cite{wu2019empirical}.
Along with this process, there is an obvious need for algorithms
and libraries on EVM. 
% Nowadays, most of Solidity programming libraries focus on tokenization,
% and many ``basic'' operations (such as finding a substring)
% force programmer to implement it himself (usually inefficiently)
% or in external libraries.
In~\cite{jabbar2020investigating} 
the authors performed an interesting analysis 
of computational costs using gas consumption as the metric.
The gas price prediction~\cite{pierro2022user, mars2021machine, kim2021predicting}
and transaction fee optimization~\cite{li2020trace, di2022profiling, khan2022gas} 
is an active subject of study.

In this paper, 
propose an efficient implementation
of several exact pattern matching algorithms
in the Solidity/YUL language,
and evaluate the gas usage of
mentioned algorithms on the Ethereum Virtual Machine.
We evaluate the algorithms' gas fee and execution time for different parameters 
(such as pattern length, alphabet size, and text size).
We show that some of those algorithms significantly reduce the gas fee and execution time compared to the existing Solidity library.
The following contributions of this work can be enumerated:
(i) We adapt and implement several exact string matching algorithms 
for Ethereum Virtual Machine. 
(ii) We present the performance of all implemented algorithms 
in the Ethereum blockchain environment.
(iii) We show the gas fee and execution time reduction comparing to popular Solidity library.

Section~\ref{sec:solution}
defines the problem of exact pattern matching
and describes all the implemented algorithms.
Section~\ref{sec:exp} presents the results of 
performed experiments in terms of gas usage.
Finally, section~\ref{sec:conclusion} 
concludes the results and suggests future work.

% \subsection{Related work}
% \label{subsec:relwork}

\section{Our Approach}
\label{sec:solution}

\subsection{Problem}
\label{subsec:problem}

Exact string matching is one of the most explored problems in computer science.
The problem can be stated as follows: For a given text $T[0 \ldots n-1]$,
and a pattern $P[0 \ldots m-1]$, $m \le n$, both over a common alphabet 
$\sum$ of size $\sigma$, 
report all 
% positions $i$ of $P$ in $T$, 
occurrences of $P$ in $T$,
such that $P[0 \ldots m-1] = T[i \ldots i + m - 1]$,
where $i \le n-m-1$.

\subsection{Algorithms}
\label{subsec:algorithms}
The string matching algorithms constitute an essential component in many software applications~\cite{charras2004handbook}.
Over the years, tens of algorithms have been invented,
most of which are modifications of the older ones~\cite{grabowski2009new, FL13}.
We adapted and implemented several classic exact text matching algorithms.
% The implementation of some of them implied changes compared to the original one
% which results from the limitations and EVM specific.
The Solidity language and EVM specification
(and limitation)
implied changes to original algorithms implementation.
We adapted and optimized the algorithms to take advantage of 32-byte word 
and reduce the number of expensive instructions 
(i.e., bitwise shifts) 
as much as possible.
Those changes are rather technical tricks that do not change the algorithm logic.
Many instructions 
(i.e., SIMD\footnote{A proposal (\href{https://eips.ethereum.org/EIPS/eip-616}{EIP-616}) 
to provide SIMD in EVM was created by Greg Colvin in 2017 but has not been implemented yet.}, AVX)
that are available in modern CPU are not available in EVM yet, 
and thus some improvements cannot be implemented.
% Additionally, some of the algorithms are 
% implemented in two variants, the standard and optimized
% (the details are given below).
% On the other hand, lack of many instructions (in EVM) which are available 
% in modern CPUs limits optimization opportunities

\subsubsection{Naive algorithm}
\label{subsubsec:naive}

The Naive (also called Brute-Force) approach to this problem 
is to scan text $T$ using a window of size $m$.
The window starts at position 0 and moves towards the end of the string
(say from left to right).
At each step, the content of the window is compared 
with the pattern character by character. 
If all $m$ characters match, the position is reported.
If there is a mismatch, the window is shifted by one position to the right.
The complexity is $O(nm)$ in the worst case, and if $\sigma \ge 2$,
then the average complexity equals $O(n)$, 
which was experimentally supported in~\cite{HS91}.

% \begin{algorithm}
% \caption{Brute-Force}\label{lst:lst_bf}
% \begin{algorithmic}[1]
% \Function{Brute\_Force}{$P[0 \ldots m - 1],T[0 \ldots n - 1]$}
% \For {$i \gets 0 \ldots n - m$}
%   \If {$T[i \ldots i + m - 1] = P[0 \ldots m - 1]$}
%     \State $\txt{report~} i$
%   \EndIf
% \EndFor
% \EndFunction
% \end{algorithmic}
% \end{algorithm}

\subsubsection{Knuth-Morris-Pratt}
\label{subsubsec:kmp}

One of the first solutions that reduce the number of character comparisons
to find a pattern in the text is Knuth-Morris-Pratt (KMP) algorithm~\cite{KMP77}.
KMP reads the pattern and builds a lookup table $N[0 \ldots m-1]$,
which contains information on how many characters can be skipped if a mismatch occurs.
The algorithm sequentially compares characters between pattern $P$ and text $T$
from left to right. Once all $m$ characters are matched, the position is reported.
If a mismatch occurs, the algorithm reads how many characters can be skipped from table $N$.
KMP compares between $n$ and $2n-1$ characters, 
the search complexity is $O(n)$ and the $N$ table is done in $O(m)$.

\subsubsection{Boyer-Moore-Horspool}
\label{subsubsec:bmh}

Boyer-Moore-Horspool (BMH) algorithm~\cite{Hor80} is a simplified 
variant of Boyer-Moore (BM)~\cite{BM77}.
The algorithm compares characters from right to left, and if a mismatch occurs, 
then the window is shifted 
according to so-called \textit{bad-character heuristic}~\cite{Hor80}.
Searching takes $O(nm)$ time in the worst case,
$O(n log_\sigma(m)/m)$ in the average case,
and $O(n/m)$ in the best case.

\subsubsection{Rabin-Karp}
\label{subsubsec:rk}

Rabin-Karp (RK)
algorithm~\cite{KR87}
is the first that uses \textit{rolling-hash} for text search purposes.
The algorithm calculates a hash for the pattern $P[0 \ldots m - 1]$
and text window $T[0 \ldots m - 1]$,
then moves the window towards the end of the text.
At each step, $i$, the hash is recalculated by adding the character that enters the window $T[i + m]$ 
and removing the one that moves outside the window $T[i]$.
The average complexity of this algorithm is $O(n + m)$, and $O(nm)$ in the worst case.

\subsubsection{Shift-Or}
\label{subsubsec:so}

Shift-Or (SO) algorithm~\cite{BYG92} 
simulates Nondeterministic Finite Automata (NFA)~\cite{rabin1959finite}.
It uses the bitwise techniques and is efficient if 
$m \le w$, where $w$ is the machine word size.
In EVM, the machine word size is 256-bit, so the algorithm 
performs the best if the pattern has at most 256 characters.
The algorithm can find a larger pattern, 
but in that case, 
it searches for the prefix (of size $w$) and verifies the reported positions.
One workaround for this limitation was presented in~\cite{SGF19}
where authors used End-Tagged Dense Code~\cite{BINPecir03} to reduce the pattern size.
The complexity of this algorithm is independent of the pattern length
and equals $O( n\lceil m/w \rceil )$, which gives $O(n)$ for $m = O(w)$.

\subsubsection{Backward Nondeterministic Dawg Matching}
\label{subsubsec:bndm}

Backward Nondeterministic Dawg Matching (BNDM)~\cite{NR98b}
is a Directed Acyclic Word Graph simulation implemented 
with bit-parallel techniques.
The algorithm, like BMH, compares the characters from the last character in the window,
and if the character does not occur in the pattern,
the window is shifted by $m - x$ 
($x$ is the length of the suffix that matches) 
characters forward.
Like SO, the max pattern length depends on the machine word size.
The complexity is $P(n/m)$ in the best case and $O(nm)$ in the worst case 
and $O(n log_\sigma m/m)$.

\subsubsection{Stringutils}
\label{subsubsec:stringutils}

StringUtils~\cite{stringutils}, 
is a popular Solidity library for string operations that most developers 
would copy into their programs and deploy along with their smart contracts~\cite{iyer2018building}.
There are several functions supported, 
but we are primarily interested in the ``find'' operation, 
which searches the first occurrence of pattern $P$
in text $T$ and returns the ``slice'' 
(a data structure representing the substring of the text). 
If the $i$ is the position of the first occurrence 
of the pattern in the text, then the returned substring is $T[i \ldots n - 1]$.
In order to evaluate its performance, 
we had to modify it, 
so it returns all the occurrences of the pattern in the text likewise other presented algorithms.

\section{Experimental results}
\label{sec:exp}

In order to evaluate the performance of the algorithms, 
we performed various experiments.
We tested the algorithms in the Ethereum network using
Ganache v6.12.2 (ganache-core: 2.13.2)\footnote{\href{https://trufflesuite.com/ganache/}{trufflesuite.com/ganache}}
(a personal blockchain for development).
The algorithms were implemented in Solidity language 
interleaved with inline assembly statements written in YUL.
All source codes were compiled with solc 
v0.8.11 compiler with optimizer enabled for 200 runs
and shared publicly on 
Github\footnote{\href{https://github.com/rsusik/pattern-matching-in-blockchain}{github.com/rsusik/pattern-matching-in-blockchain}}.
The smart contract was deployed on the Rinkeby network
% \href{https://rinkeby.etherscan.io/address/0x9Fb22d8d82FcF1c5321D5acf75eE917CF936E257}{0x9Fb22d8d82FcF1c5321D5acf75eE917CF936E257}.
\footnote{\href{https://rinkeby.etherscan.io/address/0x9Fb22d8d82FcF1c5321D5acf75eE917CF936E257}{0x9Fb22d8d82FcF1c5321D5acf75eE917CF936E257}}.
The experiments were executed on a machine equipped with 

% Intel(R) Core(TM) i7-5600U CPU 3.20GHz,
% (64 KB L1, 512KB L2 and 4MB L3 memory), 16GB of DDR3 1600 MHz RAM,
% and running under ArchLinux with Linux kernel version 5.18.7.
Intel(R) Core(TM) i5-3570 CPU 3.4 GHz,
(256~KB L1, 1~MB L2, and 6~MB L3 memory), 16 GB of DDR3 1333 MHz RAM,
running under Fedora 28 64-bit OS.
As a competing algorithm for comparison, we took a widely used and popular 
StringUtils\footnote{\href{https://github.com/Arachnid/solidity-stringutils}{github.com/Arachnid/solidity-stringutils}}
library.
We are unaware of any other fast implementation of exact string matching algorithms
% (or any other string matching algorithms) 
on the blockchain than the functions available in StringUtils.
The tests were performed on 
% substrings (1KiB, 16KiB, 128KiB) of 
datasets
(\texttt{dna}, \texttt{english}, \texttt{proteins}, \texttt{sources}) from
Pizza~\&~Chilli corpus\footnote{\href{http://pizzachili.dcc.uchile.cl}{pizzachili.dcc.uchile.cl}}.
%Tests were run on multiple datasets of different alphabet sizes
%such as:
% A multiple
% alphabet sizes were tested 
% \{\texttt{dna}, \texttt{proteins}, \texttt{english}, \texttt{sources}\},
Algorithms were tested using multiple 
patterns sizes $m \in \{$4, 8, 12, 16, 24, 32, 64, 128, 256, 512$\}$,
and text sizes $n \in \{$1KiB, 16KiB, 128KiB$\}$
(substrings of mentioned datasets).
% For each test case we search 11 random patterns and show the median value.
We generated 11 patterns for each test case and presented the median value 
(gas or execution time) of searching them.
% All the algorithms were implemented to return (by emitting an event) 
% the number of pattern occurrences in the text, 
% but they can be configured to report all pattern positions.
In the first set of analyses, 
we investigated the impact of the alphabet, 
text size, and pattern size on gas usage. 
We noticed a considerable difference in gas usage for different $m$.

\begin{figure*}[!ht]
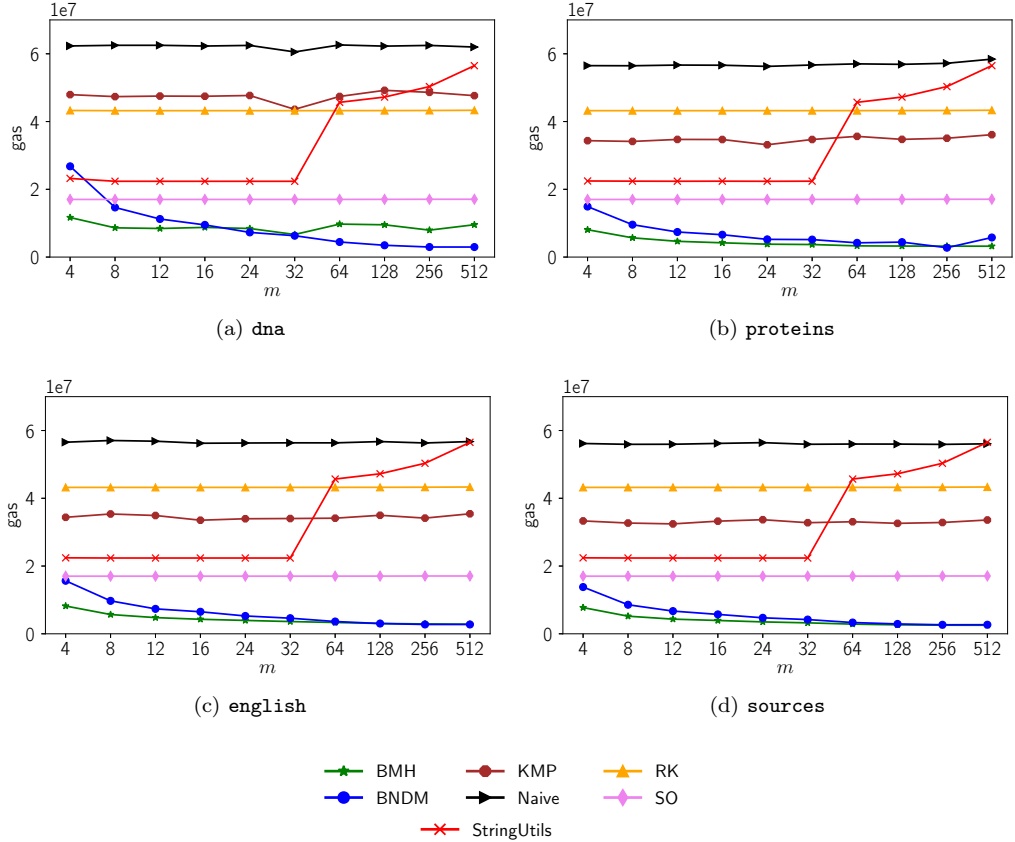

  \centering
  \subfloat[\texttt{dna}]{
      \label{fig:result_gasm_dna}
      \resizebox{0.90\columnwidth}{!}{\input{gas_m_n_128KiB__dna.pgf}}
  }
  \subfloat[\texttt{proteins}]{
      \label{fig:result_gasm_proteins}
      \resizebox{0.90\columnwidth}{!}{\input{gas_m_n_128KiB__proteins.pgf}}
  }

  \subfloat[\texttt{english}]{
      \label{fig:result_gasm_english}
      \resizebox{0.90\columnwidth}{!}{\input{gas_m_n_128KiB__english.pgf}}
  }
  \subfloat[\texttt{sources}]{
      \label{fig:result_gasm_sources}
      \resizebox{0.90\columnwidth}{!}{\input{gas_m_n_128KiB__sources.pgf}}
  }

  \subfloat{
      \label{fig:result_gasm_legend}
      \resizebox{0.70\columnwidth}{!}{\input{gas_m_n_128KiB__legend.pgf}}
  }
  \caption{Gas usage in function of pattern size for $n=128~\text{KiB}$}
  \label{fig:result_gasm}
\end{figure*}

In Fig.~\ref{fig:result_gasm} 
we can clearly see that the StringUtils function increases rapidly 
once the pattern size exceeds 32 characters.
It can be easily explained, 
the StringUtils highly depends on the fact that the machine word size in EVM is 32 bytes.
For patterns with at most 32 characters ($m \le 32$), 
the algorithm packs all the pattern characters into a 32-byte variable,
compares it against the masked text window
and finally shifts the text window by one position.
If the pattern is longer than 32 characters, 
the algorithm calculates the hash (\texttt{keccak256})
of the pattern, 
compares it against the hash of the text window,
and then shifts the text window by one.
The cost of \texttt{keccak256} depends on the length of the input,
which is why the cost grows if $m$ increases.
On the other hand,
the Boyer-Moore-Horspool algorithm takes advantage of longer patterns as it allows to 
make larger jumps 
(if the first character does not exist in the pattern, 
the algorithm skips $m$ positions of the text).
We find that the BHM wins in almost all cases, 
and only BNDM is comparable.
The most striking fact to emerge from these results is that the 
BMH reduces the gas usage by up to 22-folds comparing to StringUtils.
Interesting is the fact that the StringUtils
has an even worse result than the Naive approach 
for \texttt{sources}, and $m=512$.

\begin{table*}[!htb]
  \begin{center}
  % \resizebox{\linewidth}{!}{%
\begin{tabular}{lrrrrrrrr}
\toprule
Set         & \multicolumn{2}{c}{dna}             & \multicolumn{2}{c}{english}          & \multicolumn{2}{c}{proteins}           & \multicolumn{2}{c}{sources}   \\
            &    gas          &       fee         &     gas         &       fee          &      gas         &       fee           &     gas          &       fee  \\
Algorithm   &                 &                   &                 &                    &                  &                     &                  &            \\
\midrule
BMH         &   9.55          &   \$298.34        &    2.83         &    \$88.43         &     \textbf{3.21}&   \textbf{\$100.29} &    \textbf{2.53} &    \textbf{\$78.92} \\
BNDM        &   \textbf{2.96} &  \textbf{\$92.49} &   \textbf{2.75} &    \textbf{\$85.99}&     5.78         &   \$180.47          &    2.68          &    \$83.65 \\
KMP         &  47.65          &  \$1488.91        &   35.42         &  \$1106.95         &    36.12         &  \$1128.80          &   33.59          &  \$1049.69 \\
Naive       &  61.99          &  \$1937.23        &   56.74         &  \$1773.07         &    58.42         &  \$1825.76          &   56.07          &  \$1752.33 \\
RK          &  43.33          &  \$1353.93        &   43.33         &  \$1353.93         &    43.34         &  \$1354.31          &   43.33          &  \$1353.93 \\
SO          &  17.06          &   \$533.21        &   17.06         &   \$533.21         &    17.07         &   \$533.35          &   17.06          &   \$533.21 \\
StringUtils &  56.50          &  \$1765.74        &   56.50         &  \$1765.74         &    56.51         &  \$1765.93          &   56.50          &  \$1765.74 \\
\bottomrule
\end{tabular}

% }
\end{center}
\caption{Gas usage (in millions) and fee of searching long pattern ($m = 512$) in 128 KiB text}
\label{table:gas_m_512}
\end{table*}

% \begin{table*}[!htb]
%   \begin{center}
%   % \resizebox{\linewidth}{!}{%
% \input{tab1price.tex}
% % }
% \end{center}
% \caption{Approximate cost in USD of searching pattern ($m = 512$) in 128 KiB text. Gas price (25 Gwei) and Ethereum exchange rate (1250 USD)}
% \label{table:price_m_512}
% \end{table*}

Table\ref{table:gas_m_512}
shows gas usage and its price (fee) of searching $m=512$ 
pattern in 128KiB text.
In this case,
if we assume the current\footnote{As per 2022.07.10} 
gas price (about 25 Gwei)
and USD/ETH exchange rate (1250 USD),
the approximate cost of searching pattern of 
$m=512$ characters in \texttt{sources} dataset using StringUtils is about \$1766 
whereas the same using BMH is about \$79.
The BMH wins for \texttt{proteins} and \texttt{sources},
but in case of \texttt{dna} and \texttt{english}
the BNDM dominates.
However, only in the case of \texttt{dna} 
the difference between BMH and BNDM is notable.

\begin{table*}[!htb]
  \begin{center}
  % \resizebox{\linewidth}{!}{%
\begin{tabular}{lrrrrrrrr}
\toprule
             & Algorithm &     BMH &    BNDM &     KMP &   Naive &      RK &      SO &  StringUtils \\
Set & n &         &         &         &         &         &         &              \\
\midrule
dna & 1 KiB &  111.89 &  124.19 &  358.08 &  478.56 &  357.66 &  176.04 &       195.31 \\
    & 16 KiB &   67.97 &   74.18 &  366.65 &  473.92 &  331.00 &  132.12 &       171.79 \\
    & 128 KiB &   66.84 &   72.44 &  362.15 &  475.25 &  329.66 &  129.77 &       170.64 \\
english & 1 KiB &   77.18 &   95.52 &  290.64 &  450.87 &  357.66 &  176.04 &       195.31 \\
    & 16 KiB &   34.57 &   52.27 &  261.74 &  431.39 &  331.00 &  132.12 &       171.79 \\
    & 128 KiB &   32.70 &   49.62 &  255.79 &  429.09 &  329.66 &  129.77 &       170.64 \\
proteins & 1 KiB &   77.70 &   96.22 &  297.96 &  452.59 &  357.66 &  176.04 &       195.31 \\
    & 16 KiB &   34.30 &   49.81 &  259.82 &  430.62 &  331.00 &  132.12 &       171.79 \\
    & 128 KiB &   32.12 &   50.30 &  264.74 &  432.02 &  329.68 &  129.78 &       170.91 \\
sources & 1 KiB &   86.59 &  105.59 &  289.87 &  456.54 &  357.66 &  176.04 &       195.31 \\
    & 16 KiB &   32.03 &   48.07 &  244.69 &  425.78 &  331.00 &  132.12 &       171.79 \\
    & 128 KiB &   30.01 &   43.70 &  253.65 &  428.74 &  329.66 &  129.77 &       170.64 \\
\bottomrule
\end{tabular}

% }
\end{center}
\caption{Gas usage per text character of searching short ($m=16$) pattern}
\label{table:gas_16}
\end{table*}

Table~\ref{table:gas_16} presents gas usage per text character.
All the algorithms, despite Naive, 
are more expensive for very short texts (1 KiB)
than for longer ones (16 KiB and 128 KiB).
The gas usage per character falls with the text size increase.
The difference between 1 KiB and 16 KiB is more considerable than between 16 KiB and 128 KiB.
We see, in this case, 
the impact of function initial steps and the preprocessing phase on transaction cost. 
Both take the same amount of resources if the pattern size is fixed.
It is also the reason why it only slightly affects the Naive algorithm, 
which does not have preprocessing phase.
% It can be noticed that there is little difference 
% between 16 KiB and 128 KiB
% when increasing the text size.
% (assuming the same patterns size)
% which highly depends on pattern size.

\begin{figure*}[!ht]
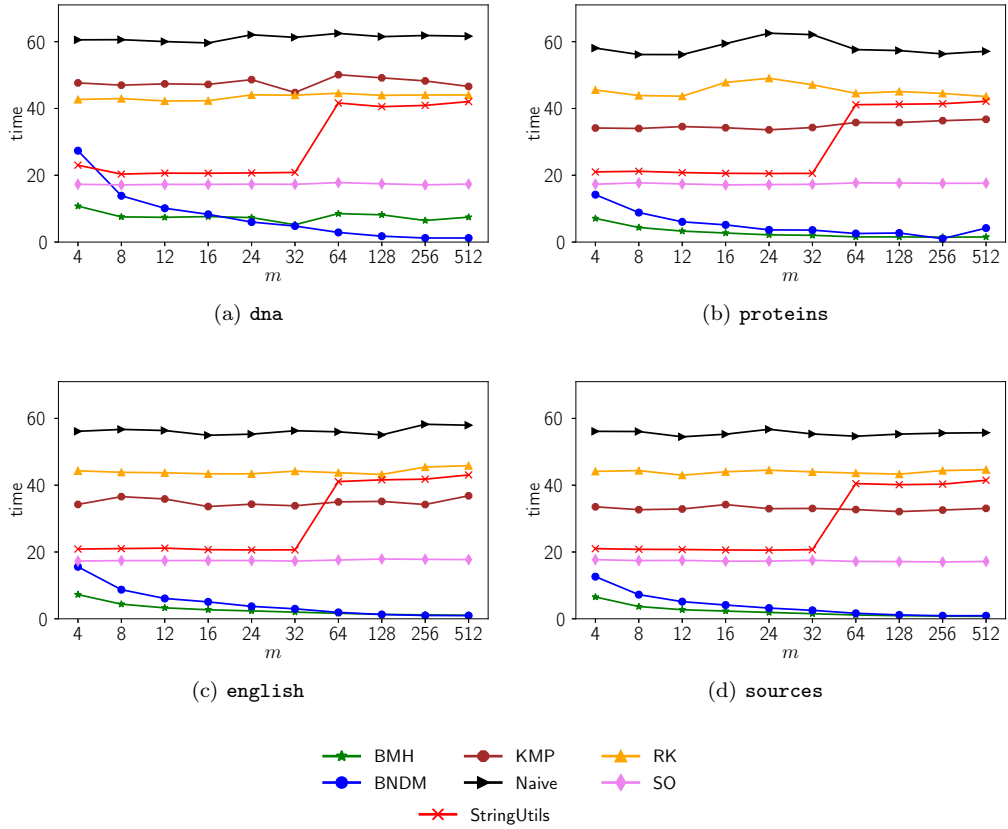

  \centering
  \subfloat[\texttt{dna}]{
      \label{fig:result_timem_dna}
      \resizebox{0.90\columnwidth}{!}{\input{time_m_n_128KiB__dna.pgf}}
  }
  \subfloat[\texttt{proteins}]{
      \label{fig:result_timem_proteins}
      \resizebox{0.90\columnwidth}{!}{\input{time_m_n_128KiB__proteins.pgf}}
  }

  \subfloat[\texttt{english}]{
      \label{fig:result_timem_english}
      \resizebox{0.90\columnwidth}{!}{\input{time_m_n_128KiB__english.pgf}}
  }
  \subfloat[\texttt{sources}]{
      \label{fig:result_timem_sources}
      \resizebox{0.90\columnwidth}{!}{\input{time_m_n_128KiB__sources.pgf}}
  }

  \subfloat{
      \label{fig:result_timem_legend}
      \resizebox{0.70\columnwidth}{!}{\input{time_m_n_128KiB__legend.pgf}}
  }
  \caption{Time (in seconds) in function of pattern size for $n=128~\text{KiB}$}
  \label{fig:result_timem}
\end{figure*}

In addition to the gas usage, 
we measured searching time. 
Figure~\ref{fig:result_timem}
presents median time (in seconds)
of searching patterns in 128 KiB text.
As expected, 
the time and gas are correlated. 
However, we noticed some discrepancies between them. 
% The discrepancy grows with the pattern length ($m$).
For instance, 
in the case of \texttt{sources} dataset, 
the StringUtils needs 22 times more gas than BMH to find a pattern of $m=512$ characters, 
whereas it needs 55-fold more time, 
which means the result in terms of time is 150\% higher than for the gas usage.
% which gives about 150\% better result for time.
% which means the BMH has 150\% better result than StringUtils
% in terms of time over .
Nevertheless, 
the difference is much smaller for very short patterns 
($m=4$, and the same other parameters).
The StringUtils time and the gas usage are 3.22 and 2.90 higher than BMH (respectively),
resulting in about 11\% more in time than gas consumption.
% % We noticed the greatest discrepancy %
% % for \texttt{dna} and $m=512$ between BNDM and StringUtils.
% % In this case, 
% % BNDM consumes 19 times less gas than StringUtils
% % and is 31-fold faster, and thus, 
% % the performance in function of the time is 63\% better than for the gas usage.

\begin{figure*}[!ht]
  \centering
  \resizebox{1.0\columnwidth}{!}{\input{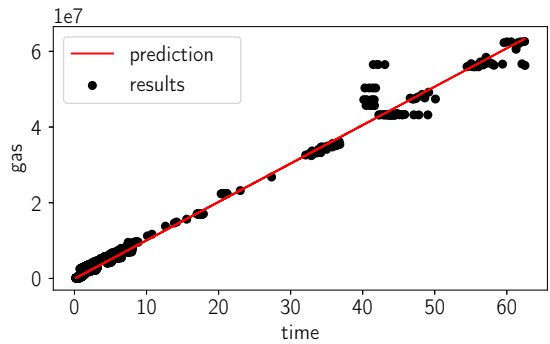}}
  \caption{Gas usage in function of execution time (in seconds)}
  \label{fig:result_gas_vs_time}
\end{figure*}

Finally, 
we accumulated all the results 
and displayed them in a single chart.
Figure~\ref{fig:result_gas_vs_time}
presents the gas usage in the function of the execution time
along with the trend line.
% This chart is interesting in several ways.
We can clearly see that the execution time and the gas usage are correlated.
The coefficients of linear regression allow estimating the cost of code execution approximately.
We found that one second of code execution 
(assuming our environment specification) 
would cost about \$31.71.

\section{Conclusion}
\label{sec:conclusion}

In this work, 
we adapted exact pattern matching algorithms to EVM architecture,
implemented the algorithms in Solidity language combined with YUL assembly,
and performed extensive tests using the Ethereum blockchain.
We empirically proved that the gas usage could be significantly reduced in all the test cases.
The experiments confirmed the technical (smaller computational time)
and financial (smaller costs) advantages of the proposed approach.
We demonstrated that the cost of searching patterns could be reduced by up to 22 times 
(\$78.92 vs \$1765.74)
and the execution time by up to 55 times (41.47s. vs. 0.75s.)
when compared to StringUtils.% (popular open source Solidity library).

\section*{Acknowledgement}
We thank Sz. Grabowski (\textit{Lodz University of~Technology}) 
for helpful discussions and his valuable suggestions.

\bibliographystyle{abbrv}
\bibliography{main}

\begin{thebibliography}{10}

\bibitem{stringutils}
Stringutils.
\newblock \url{https://github.com/Arachnid/solidity-stringutils}.
\newblock Accessed: 2022-04-15.

\bibitem{adere2022blockchain}
E.~M. Adere.
\newblock Blockchain in healthcare and iot: A systematic literature review.
\newblock {\em Array}, page 100139, 2022.

\bibitem{almasoud2020smart}
A.~S. Almasoud, F.~K. Hussain, and O.~K. Hussain.
\newblock Smart contracts for blockchain-based reputation systems: A systematic
  literature review.
\newblock {\em Journal of Network and Computer Applications}, 170:102814, 2020.

\bibitem{amler2021defi}
H.~Amler, L.~Eckey, S.~Faust, M.~Kaiser, P.~Sandner, and B.~Schlosser.
\newblock Defi-ning defi: Challenges \& pathway.
\newblock In {\em 2021 3rd Conference on Blockchain Research \& Applications
  for Innovative Networks and Services (BRAINS)}, pages 181--184. IEEE, 2021.

\bibitem{BYG92}
R.~A. Baeza-Yates and G.~H. Gonnet.
\newblock A new approach to text searching.
\newblock {\em Communications of the {ACM}}, 35(10):74--82, 1992.

\bibitem{belchior2021survey}
R.~Belchior, A.~Vasconcelos, S.~Guerreiro, and M.~Correia.
\newblock A survey on blockchain interoperability: Past, present, and future
  trends.
\newblock {\em ACM Computing Surveys (CSUR)}, 54(8):1--41, 2021.

\bibitem{BM77}
R.~S. Boyer and J.~S. Moore.
\newblock A fast string searching algorithm.
\newblock {\em Communications of the {ACM}}, 20(10):762--772, 1977.

\bibitem{BINPecir03}
N.~Brisaboa, E.~Iglesias, G.~Navarro, and J.~L. Param\'a.
\newblock An efficient compression code for text databases.
\newblock In {\em Proc. 25th European Conference on Information Retrieval
  Research (ECIR'03)}, LNCS 2633, pages 468--481. Springer, 2003.

\bibitem{Buterin2013}
V.~Buterin.
\newblock Ethereum white paper: A next generation smart contract \&
  decentralized application platform.
\newblock 2013.

\bibitem{charras2004handbook}
C.~Charras and T.~Lecroq.
\newblock {\em Handbook of exact string matching algorithms}.
\newblock Citeseer, 2004.

\bibitem{di2022profiling}
A.~Di~Sorbo, S.~Laudanna, A.~Vacca, C.~A. Visaggio, and G.~Canfora.
\newblock Profiling gas consumption in solidity smart contracts.
\newblock {\em Journal of Systems and Software}, 186:111193, 2022.

\bibitem{FL13}
S.~Faro and T.~Lecroq.
\newblock The exact online string matching problem: A review of the most recent
  results.
\newblock {\em ACM Computing Surveys (CSUR)}, 45(2), mar 2013.

\bibitem{gad2022emerging}
A.~G. Gad, D.~T. Mosa, L.~Abualigah, and A.~A. Abohany.
\newblock Emerging trends in blockchain technology and applications: A review
  and outlook.
\newblock {\em Journal of King Saud University-Computer and Information
  Sciences}, 2022.

\bibitem{grabowski2009new}
S.~Grabowski.
\newblock New algorithms for exact and approximate text matching.
\newblock online.

\bibitem{guo2022survey}
H.~Guo and X.~Yu.
\newblock A survey on blockchain technology and its security.
\newblock {\em Blockchain: Research and Applications}, page 100067, 2022.

\bibitem{Hor80}
R.~N. Horspool.
\newblock Practical fast searching in strings.
\newblock {\em Software: Practice and Experience}, 10(6):501--506, 1980.

\bibitem{HS91}
A.~Hume and D.~Sunday.
\newblock Fast string searching.
\newblock {\em Software: Practice and Experience}, 21(11):1221--1248, 1991.

\bibitem{iyer2018building}
K.~Iyer and C.~Dannen.
\newblock {\em Building Games with Ethereum Smart Contracts}.
\newblock Springer, 2018.

\bibitem{jabbar2020investigating}
A.~Jabbar and S.~Dani.
\newblock Investigating the link between transaction and computational costs in
  a blockchain environment.
\newblock {\em International Journal of Production Research},
  58(11):3423--3436, 2020.

\bibitem{KR87}
R.~M. Karp and M.~O. Rabin.
\newblock Efficient randomized pattern-matching algorithms.
\newblock {\em {IBM} Journal of Research and Development}, 31(2):249--260,
  1987.

\bibitem{khan2022gas}
M.~M.~A. Khan, H.~M.~A. Sarwar, and M.~Awais.
\newblock Gas consumption analysis of ethereum blockchain transactions.
\newblock {\em Concurrency and Computation: Practice and Experience},
  34(4):e6679, 2022.

\bibitem{kim2021predicting}
H.-M. Kim, G.-W. Bock, and G.~Lee.
\newblock Predicting ethereum prices with machine learning based on blockchain
  information.
\newblock {\em Expert Systems with Applications}, 184:115480, 2021.

\bibitem{king2012ppcoin}
S.~King and S.~Nadal.
\newblock Ppcoin: Peer-to-peer crypto-currency with proof-of-stake.
\newblock {\em self-published paper, August}, 19(1), 2012.

\bibitem{KMP77}
D.~E. Knuth, J.~H. Morris, and V.~R. Pratt.
\newblock Fast pattern matching in strings.
\newblock {\em {SIAM} Journal on Computing}, 6(1):323--350, 1977.

\bibitem{kumar2020novel}
A.~Kumar, R.~Krishnamurthi, A.~Nayyar, K.~Sharma, V.~Grover, and E.~Hossain.
\newblock A novel smart healthcare design, simulation, and implementation using
  healthcare 4.0 processes.
\newblock {\em IEEE Access}, 8:118433--118471, 2020.

\bibitem{li2020trace}
C.~Li, S.~Nie, Y.~Cao, Y.~Yu, and Z.~Hu.
\newblock Trace-based dynamic gas estimation of loops in smart contracts.
\newblock {\em IEEE Open Journal of the Computer Society}, 1:295--306, 2020.

\bibitem{li2021decentralized}
Y.~Li, J.~Wei, J.~Yuan, Q.~Xu, and C.~He.
\newblock A decentralized music copyright operation management system based on
  blockchain technology.
\newblock {\em Procedia Computer Science}, 187:458--463, 2021.

\bibitem{marchesi2020design}
L.~Marchesi, M.~Marchesi, G.~Destefanis, G.~Barabino, and D.~Tigano.
\newblock Design patterns for gas optimization in ethereum.
\newblock In {\em 2020 IEEE International Workshop on Blockchain Oriented
  Software Engineering (IWBOSE)}, pages 9--15. IEEE, 2020.

\bibitem{mars2021machine}
R.~Mars, A.~Abid, S.~Cheikhrouhou, and S.~Kallel.
\newblock A machine learning approach for gas price prediction in ethereum
  blockchain.
\newblock In {\em 2021 IEEE 45th Annual Computers, Software, and Applications
  Conference (COMPSAC)}, pages 156--165. IEEE, 2021.

\bibitem{mollah2020blockchain}
M.~B. Mollah, J.~Zhao, D.~Niyato, Y.~L. Guan, C.~Yuen, S.~Sun, K.-Y. Lam, and
  L.~H. Koh.
\newblock Blockchain for the internet of vehicles towards intelligent
  transportation systems: A survey.
\newblock {\em IEEE Internet of Things Journal}, 8(6):4157--4185, 2020.

\bibitem{nakamoto2008bitcoin}
S.~Nakamoto.
\newblock Bitcoin: A peer-to-peer electronic cash system.
\newblock {\em Decentralized Business Review}, page 21260, 2008.

\bibitem{NR98b}
G.~Navarro and M.~Raffinot.
\newblock A bit-parallel approach to suffix automata: Fast extended string
  matching.
\newblock In {\em Proc. Annual Symposium on Combinatorial Pattern Matching
  ({CPM})}, pages 14--33. Springer, 1998.

\bibitem{nizamuddin2019decentralized}
N.~Nizamuddin, K.~Salah, M.~A. Azad, J.~Arshad, and M.~Rehman.
\newblock Decentralized document version control using ethereum blockchain and
  ipfs.
\newblock {\em Computers \& Electrical Engineering}, 76:183--197, 2019.

\bibitem{pierro2022user}
G.~A. Pierro, H.~Rocha, S.~Ducasse, M.~Marchesi, and R.~Tonelli.
\newblock A user-oriented model for oracles’ gas price prediction.
\newblock {\em Future Generation Computer Systems}, 128:142--157, 2022.

\bibitem{rabin1959finite}
M.~O. Rabin and D.~Scott.
\newblock Finite automata and their decision problems.
\newblock {\em IBM journal of research and development}, 3(2):114--125, 1959.

\bibitem{sober2021voting}
M.~Sober, G.~Scaffino, C.~Spanring, and S.~Schulte.
\newblock A voting-based blockchain interoperability oracle.
\newblock In {\em 2021 IEEE International Conference on Blockchain
  (Blockchain)}, pages 160--169. IEEE, 2021.

\bibitem{SGF19}
R.~Susik, S.~Grabowski, and K.~Fredriksson.
\newblock Revisiting multiple pattern matching.
\newblock {\em Computing and Informatics}, 38(4):937--962, 2019.

\bibitem{szabo1996smart}
N.~Szabo.
\newblock Smart contracts: building blocks for digital markets.
\newblock {\em EXTROPY: The Journal of Transhumanist Thought,(16)}, 18(2):28,
  1996.

\bibitem{turing1936computable}
A.~M. Turing.
\newblock On computable numbers, with an application to the
  entscheidungsproblem.
\newblock {\em J. of Math}, 58(345-363):5, 1936.

\bibitem{wu2019empirical}
K.~Wu.
\newblock An empirical study of blockchain-based decentralized applications.
\newblock {\em arXiv preprint arXiv:1902.04969}, 2019.

\bibitem{zhang2022blockchain}
Y.~Zhang, L.~Chen, M.~Battino, M.~A. Farag, J.~Xiao, J.~Simal-Gandara, H.~Gao,
  and W.~Jiang.
\newblock Blockchain: An emerging novel technology to upgrade the current fresh
  fruit supply chain.
\newblock {\em Trends in Food Science \& Technology}, 2022.

\end{thebibliography}

\appendix

\section*{Appendix A}
In this section we provide complementary results.

\begin{table*}[!htb]
  \begin{center}
  % \resizebox{\linewidth}{!}{%
\begin{tabular}{llrrrrrrrrrr}
\toprule
             & m &    4   &    8   &    12  &    16  &    24  &    32  &    64  &    128 &    256 &    512 \\
Set & Algorithm &        &        &        &        &        &        &        &        &        &        \\
\midrule
dna & BMH &  10.76 &   7.54 &   7.41 &   7.62 &   7.34 &   5.19 &   8.52 &   8.16 &   6.46 &   7.45 \\
             & BNDM &  27.36 &  13.86 &  10.13 &   8.30 &   5.99 &   4.80 &   2.88 &   1.76 &   1.22 &   1.21 \\
             & KMP &  47.65 &  46.97 &  47.37 &  47.21 &  48.62 &  44.76 &  50.10 &  49.16 &  48.25 &  46.59 \\
             & Naive &  60.54 &  60.59 &  60.03 &  59.62 &  62.06 &  61.29 &  62.48 &  61.50 &  61.85 &  61.63 \\
             & RK &  42.69 &  42.96 &  42.24 &  42.31 &  44.07 &  43.98 &  44.56 &  43.93 &  44.04 &  44.04 \\
             & SO &  17.30 &  17.12 &  17.26 &  17.27 &  17.32 &  17.29 &  17.81 &  17.46 &  17.12 &  17.37 \\
             & StringUtils &  23.01 &  20.35 &  20.64 &  20.60 &  20.70 &  20.84 &  41.63 &  40.53 &  40.92 &  42.05 \\
english & BMH &   7.25 &   4.38 &   3.27 &   2.71 &   2.36 &   1.99 &   1.65 &   1.32 &   1.17 &   1.05 \\
             & BNDM &  15.57 &   8.72 &   6.08 &   5.07 &   3.72 &   2.99 &   1.92 &   1.29 &   0.99 &   0.95 \\
             & KMP &  34.24 &  36.57 &  35.87 &  33.60 &  34.30 &  33.83 &  35.00 &  35.15 &  34.21 &  36.83 \\
             & Naive &  56.13 &  56.70 &  56.37 &  54.94 &  55.28 &  56.30 &  55.97 &  55.07 &  58.23 &  57.95 \\
             & RK &  44.29 &  43.87 &  43.71 &  43.40 &  43.40 &  44.20 &  43.73 &  43.22 &  45.42 &  45.84 \\
             & SO &  17.30 &  17.41 &  17.43 &  17.45 &  17.44 &  17.28 &  17.60 &  17.89 &  17.80 &  17.72 \\
             & StringUtils &  20.88 &  21.02 &  21.16 &  20.70 &  20.62 &  20.64 &  41.08 &  41.60 &  41.80 &  43.07 \\
proteins & BMH &   7.08 &   4.37 &   3.28 &   2.71 &   2.16 &   2.02 &   1.55 &   1.54 &   1.47 &   1.53 \\
             & BNDM &  14.19 &   8.84 &   6.06 &   5.15 &   3.65 &   3.59 &   2.56 &   2.71 &   1.01 &   4.18 \\
             & KMP &  34.15 &  34.01 &  34.57 &  34.23 &  33.60 &  34.29 &  35.80 &  35.78 &  36.35 &  36.75 \\
             & Naive &  58.07 &  56.15 &  56.13 &  59.41 &  62.52 &  62.10 &  57.62 &  57.33 &  56.33 &  57.12 \\
             & RK &  45.54 &  43.86 &  43.68 &  47.79 &  49.04 &  47.09 &  44.52 &  45.07 &  44.50 &  43.58 \\
             & SO &  17.32 &  17.74 &  17.41 &  17.12 &  17.20 &  17.30 &  17.73 &  17.68 &  17.59 &  17.61 \\
             & StringUtils &  21.01 &  21.19 &  20.81 &  20.58 &  20.53 &  20.57 &  41.12 &  41.26 &  41.41 &  42.13 \\
sources & BMH &   6.52 &   3.67 &   2.71 &   2.31 &   1.90 &   1.53 &   1.14 &   0.93 &   0.80 &   0.75 \\
             & BNDM &  12.62 &   7.23 &   5.13 &   4.12 &   3.21 &   2.54 &   1.65 &   1.16 &   0.92 &   0.92 \\
             & KMP &  33.53 &  32.66 &  32.87 &  34.19 &  32.97 &  33.04 &  32.70 &  32.09 &  32.56 &  33.05 \\
             & Naive &  56.12 &  56.06 &  54.49 &  55.27 &  56.73 &  55.34 &  54.65 &  55.29 &  55.59 &  55.71 \\
             & RK &  44.13 &  44.38 &  42.98 &  44.01 &  44.50 &  43.97 &  43.62 &  43.31 &  44.37 &  44.63 \\
             & SO &  17.70 &  17.42 &  17.49 &  17.25 &  17.26 &  17.51 &  17.18 &  17.13 &  17.01 &  17.17 \\
             & StringUtils &  21.01 &  20.79 &  20.75 &  20.60 &  20.55 &  20.70 &  40.48 &  40.15 &  40.31 &  41.47 \\
\bottomrule
\end{tabular}

% }
\end{center}
\caption{Time (in seconds) of searching pattern in 128 KiB text}
\label{table:time_128kib}
\end{table*}

\begin{table*}[!htb]
  \begin{center}
  % \resizebox{\linewidth}{!}{%
\begin{tabular}{llrrrrrrrrrr}
\toprule
             & m &    4   &    8   &    12  &    16  &    24  &    32  &    64  &    128 &    256 &    512 \\
Set & Algorithm &        &        &        &        &        &        &        &        &        &        \\
\midrule
dna & BMH &  11.68 &   8.63 &   8.44 &   8.76 &   8.43 &   6.67 &   9.73 &   9.53 &   7.95 &   9.55 \\
             & BNDM &  26.78 &  14.64 &  11.24 &   9.49 &   7.28 &   6.30 &   4.45 &   3.48 &   2.97 &   2.96 \\
             & KMP &  47.95 &  47.36 &  47.52 &  47.47 &  47.69 &  43.61 &  47.41 &  49.22 &  48.62 &  47.65 \\
             & Naive &  62.32 &  62.50 &  62.51 &  62.29 &  62.47 &  60.55 &  62.61 &  62.26 &  62.47 &  61.99 \\
             & RK &  43.27 &  43.21 &  43.21 &  43.21 &  43.21 &  43.21 &  43.22 &  43.24 &  43.27 &  43.33 \\
             & SO &  17.02 &  17.01 &  17.01 &  17.01 &  17.01 &  17.01 &  17.02 &  17.03 &  17.06 &  17.06 \\
             & StringUtils &  23.21 &  22.37 &  22.37 &  22.37 &  22.36 &  22.36 &  45.68 &  47.23 &  50.33 &  56.50 \\
english & BMH &   8.20 &   5.69 &   4.78 &   4.29 &   3.94 &   3.61 &   3.34 &   3.04 &   2.91 &   2.83 \\
             & BNDM &  15.65 &   9.72 &   7.36 &   6.50 &   5.26 &   4.61 &   3.62 &   3.03 &   2.76 &   2.75 \\
             & KMP &  34.39 &  35.37 &  34.92 &  33.53 &  33.96 &  34.02 &  34.12 &  34.98 &  34.15 &  35.42 \\
             & Naive &  56.55 &  57.06 &  56.84 &  56.24 &  56.31 &  56.37 &  56.36 &  56.73 &  56.32 &  56.74 \\
             & RK &  43.21 &  43.21 &  43.21 &  43.21 &  43.21 &  43.21 &  43.22 &  43.24 &  43.27 &  43.33 \\
             & SO &  17.01 &  17.01 &  17.01 &  17.01 &  17.01 &  17.01 &  17.02 &  17.03 &  17.06 &  17.06 \\
             & StringUtils &  22.43 &  22.37 &  22.37 &  22.37 &  22.36 &  22.36 &  45.68 &  47.23 &  50.33 &  56.50 \\
proteins & BMH &   8.08 &   5.68 &   4.65 &   4.21 &   3.80 &   3.71 &   3.30 &   3.25 &   3.23 &   3.21 \\
             & BNDM &  14.91 &   9.56 &   7.40 &   6.59 &   5.23 &   5.16 &   4.20 &   4.42 &   2.76 &   5.78 \\
             & KMP &  34.36 &  34.13 &  34.72 &  34.70 &  33.15 &  34.70 &  35.63 &  34.74 &  35.07 &  36.12 \\
             & Naive &  56.50 &  56.47 &  56.67 &  56.63 &  56.28 &  56.70 &  57.03 &  56.89 &  57.20 &  58.42 \\
             & RK &  43.21 &  43.21 &  43.21 &  43.21 &  43.21 &  43.22 &  43.22 &  43.24 &  43.27 &  43.34 \\
             & SO &  17.01 &  17.01 &  17.01 &  17.01 &  17.01 &  17.01 &  17.02 &  17.03 &  17.06 &  17.07 \\
             & StringUtils &  22.46 &  22.40 &  22.38 &  22.40 &  22.37 &  22.40 &  45.68 &  47.23 &  50.33 &  56.51 \\
sources & BMH &   7.74 &   5.20 &   4.33 &   3.93 &   3.51 &   3.23 &   2.86 &   2.64 &   2.54 &   2.53 \\
             & BNDM &  13.80 &   8.57 &   6.70 &   5.73 &   4.72 &   4.20 &   3.33 &   2.91 &   2.66 &   2.68 \\
             & KMP &  33.32 &  32.69 &  32.44 &  33.25 &  33.68 &  32.79 &  33.07 &  32.60 &  32.85 &  33.59 \\
             & Naive &  56.17 &  55.92 &  55.95 &  56.20 &  56.42 &  55.94 &  56.01 &  56.00 &  55.91 &  56.07 \\
             & RK &  43.21 &  43.21 &  43.21 &  43.21 &  43.21 &  43.21 &  43.22 &  43.24 &  43.27 &  43.33 \\
             & SO &  17.01 &  17.01 &  17.01 &  17.01 &  17.01 &  17.01 &  17.02 &  17.03 &  17.06 &  17.06 \\
             & StringUtils &  22.42 &  22.37 &  22.37 &  22.37 &  22.36 &  22.36 &  45.68 &  47.23 &  50.33 &  56.50 \\
\bottomrule
\end{tabular}

% }
\end{center}
\caption{Gas usage (in millions) of searching pattern in 128 KiB text}
\label{table:gas_128kib}
\end{table*}

\end{document}